# Thickness-tuned transition of band topology in ZrTe$_5$ nanosheets

Jianwei Lu[1,2,*], Guolin Zheng[1,2,*], Xiangde Zhu[1] Wei Ning[1,3,†], Hongwei Zhang[1,2], YuyanHan[1], Jiyong Yang[1], Haifeng Du[1,3], Kun Yang[4], Haizhou Lu[5], Yuheng Zhang[1,2,3], and Mingliang Tian[1,2,3,†]

[1]*High Magnetic Field Laboratory, Chinese Academy of Sciences, Hefei 230031, Anhui, the People's Republic of China*

[2]*Department of physics, University of Science and Technology of China, Hefei 230026, the People's Republic of China*

[3]*Collaborative Innovation Center of Advanced Microstructures, Nanjing University, Nanjing 210093, the People's Republic of China*

[4]*National High Magnetic Field Laboratory, Florida State University, Tallahassee, Florida 32306-4005, USA*

[5]*Department of Physics, South University of Science and Technology of China, Shenzhen, China*

[*] Those authors contribute equally to this work.

[§]To whom correspondence should be addressed. E-mail: ningwei@hmfl.ac.cn (W. N.), tianml@hmfl.ac.cn (M. T.).

## Abstract

We report thickness-tuned electrical transport and Hall resistivity in highly anisotropic three-dimensional Dirac semimetal ZrTe$_5$ nanosheets. We find that when the thickness of the nanosheet is blow about 40 nm, the system takes a clear transition from topological semimetal with two bands carriers to a single band with conventional hole carriers. The resistivity peak temperature T$^*$ decreases systematically with decreasing thickness down to about 40 nm, then shifts up with the further decrease of the thickness. Analysis of the data below 40 nm indicates that the hole carriers completely dominate the transport in the entire temperature range, regardless of the temperature being below or above T$^*$. By further tracking the carrier density, we find that the Fermi level shifts consecutively downward from conduction band to the valence band as decreasing the thickness. Our experiments unambiguously reveal a highly thickness-tuned transition of band topology in ZrTe$_5$ nanosheets.

Zirconium pentatelluride ZrTe$_5$, a fascinating new three-dimensional (3D) Dirac semimetal, has attracted considerable attention recently. It hosts not only rich exotic quantum phenomena related to the chiral Fermions in its highly anisotropic three-dimensional Dirac bands (1-3), but also its electronic structure, compared with other Dirac semimetals, such as Cd$_3$As$_2$, Na$_3$Bi (4-6), presents extremely sensitivity to external perturbations such as magnetic fields, temperature, elastic tension or pressure (7-10). For example, the temperature T, according to the recent angle-resolved photoemission spectroscopy (ARPES) experiment (11), can induce a Lifshitz-type transition of electronic states from the hole band to electron band, leading to a resistance peak near the critical temperature T$^*$. Indeed, the transport measurements on bulk ZrTe$_5$ have demonstrated clearly the changes from hole-dominated states above T$^*$ to electron dominated states below T$^*$ (9, 12). While the gapless topological Dirac semimetal phase has been demonstrated in bulk ZrTe$_5$ (1-3,13,14), the recent scanning tunneling microscopy (STM) surprisingly detected a bulk band gap with topological edge states at the surface step edge and, thus, indicated that single layered ZrTe$_5$ might be a two-dimensional topological insulator (15-17), which could host the quantum spin Hall effect (QSHE) (18,19). These contrast results indicated that the thickness, as an alternative way, may effectively tune the electronic structure in ZrTe$_5$, though the mechanism has not been fully explored so far.

In this letter, we study the transport properties in ZrTe$_5$ nanosheets with thickness down to 10 nm. We find that the Lifshitz transition temperature T$^*$ systematically shifts toward low temperatures as the thickness of the nanosheets decreases down to 40 nm, indicating the suppression of the electron carriers in the Dirac band. However, when the thickness is below 40 nm, a broad resistive peak shows up at high temperatures and moves up with further decrease of the thickness. Both longitudinal resistivity and Hall resistivity measurements demonstrate that the resistance peak in thinner nanosheets (less than 40 nm) has a completely different origin from that in bulk, where carriers in a secondary hole band take over when the thickness is below about 40 nm. By tracking the carrier density, a consecutive downward shifting of the Fermi level from conduction band to valence band is demonstrated as decreasing the thickness down to 10 nm, which reveals a highly thickness-tuned transition of the band topology in ZrTe$_5$ nanosheets.

ZrTe$_5$ single crystals were grown via the iodine vapor transport method in a two-zone furnace with elements Zr (99.99%) and Te (99.99%) as described in Ref. (20). In this letter, all nanosheets were mechanically exfoliated from the bulk crystals, followed by directly transferring on Si/SiO$_2$ substrate. Hall bar devices were fabricated by standard electron-beam lithography followed by Au (80 nm)/Ti(10 nm) evaporation and lift-off process. The thickness (t) of the nanosheets were determined by side scanning of scanning electron microscopy (SEM) under high resolution mode within an uncertainty of ~5 nm. Transport measurements were performed with physical properties measurement system.

Fig. 1A shows the temperature dependence of resistivity of ZrTe$_5$ nanosheets with different thicknesses. In our thick samples (e.g. the thickness, t=160 nm), the resistance anomalous temperature $T^*$, defined from the peak position of the resistance-temperature (R-T) curve, is about 145 K, which is consistent with the $T^*$ in bulk ZrTe$_5$ single crystal (12, 21, 22). We find that $T^*$ is quite sensitive to the sample's thickness. It decreases systematically from 145 K to about 100 K as the thickness decreases from 160 nm down to 40 nm. Most strikingly, when the thickness of the sample is below 40 nm, a broad peak appears near 170 K for 33 nm thick sample, and then shifts up beyond 300 K for 20 nm thick sample. To have a better understanding of the thickness-dependent transport properties of our ZrTe$_5$ nanosheets, we have tracked the Hall resistivity of these nanosheets with different thicknesses at 2 K as shown in Fig. 1B. As expected, the Hall resistivity in 160 nm thick sample exhibits a typical multiband property and the electron band dominates the transport at 2 K. Upon decreasing the thickness to 80 nm, the Hall coefficient, $R_H \propto R_{xy}/B$ with $R_{xy}$ the Hall resistivity and B the magnetic field, shows a sign reversal from negative in low field range to positive near 4 T. The critical magnetic field, where the Hall resistivity shows a sign reversal, shifts to low field region and finally disappears in samples with thickness below 40 nm, indicating that the electron band is suppressed gradually with decreasing the thickness, but the hole carriers take over when the thickness is below 40 nm. This is consistent with the observation of the $T^*$ shifting towards low temperatures with decreasing the thickness. And the linear behavior of the Hall resistivity is also observed in a wide field range with a positive slope for nanosheets with t=25, 20, and 10 nm, as shown in Fig. 1C respectively. Hence, the role of decreasing thickness is similar to the effect of lifting up temperature, i.e., lowering down the Fermi level or shifts up the holes band. However, we should note that the nanosheets with t<40 nm present an excellent metallic behavior, which is in contrast to the semiconductive hole band above $T^*$ in thick samples. In other words, the hole carriers of the nanosheets with t<40 nm is probably not from the same hole band as indicated at high temperatures.

This longstanding question for the mechanism of the resistance peak at T$^*$ is currently attributed to the upward shift of the Fermi level from the Dirac hole band to electron band with decreasing temperature (23), i.e., the temperature-induced Lifshitz transition (11). This can also be verified by the temperature dependent Hall resistivity measurement for our nanosheets, for example, t= 55 nm, as shown in Fig. 2A. As we can see, the Hall coefficient $R_H$ near zero field exhibits a sign reversal from positive to negative near T$^*$, where the resistance exhibits a peak. Such a resistive peak originates from the temperature-induced competition between electron and hole carriers (or the shift of Fermi energy from hole band to electron band). To explore the electronic mechanism of the resistance anomaly in nanosheets with t<40 nm, we have systematically analyzed the Hall resistivity of the thinner nanosheet with t=33 nm at different temperatures. As we can see in Fig. 2B, all of the Hall resistivity shows a linear behavior with a positive slope, regardless of the temperature being below or above the peak temperature T$^*$. These results unambiguously provide an evidence that hole carriers band dominates the transport in thinner nanosheets (t<40 nm), there is not any transition of band topology through T$^*$. In other words, the origin of the broad peak presented in the R-T curves for thinner nanosheets is completely different from that observed in thick samples (t>40 nm), which cannot be attributed to the competition between electron and hole pockets.

To have a comprehensive insight on this conventional resistive anomaly in our thinner samples, we have tracked the carrier density as well as the mobility as the function of temperatures in 33 nm and 25 nm thick nanosheets, as shown in Fig. 3A and 3B. Since the Hall resistivity exhibits a linear behavior, we can evaluate the carrier density by the formula $n = 1/R_H e$. Also, the mobility μ can be calculated by the relationship $μ = σ/ne$, with $σ$ the zero field conductivity and $n$ the carrier density. As we can see, the carrier density decreases sharply with the decrease of temperature while the mobility increases gradually due to the suppression of the thermal excitation. As a result, the broad resistivity peak is likely formed due to the opposite contributions of the decreased charge carrier density n and the increased mobility μ to the electrical resistivity. This kind of resistance peak is a common feature for semimetal such as in Bi nanowire (24, 25). We note that an upward shifting of the resistance anomaly temperature in ZrTe$_5$ nanosheets below 65 nm is also found recently (26). Unfortunately, due to the lack of the Hall data, the underlying mechanism of the band carriers is not fully understood previously.

For a full understanding of the thickness-tuned band topology, we have tracked the carrier density of ZrTe$_5$ nanosheets with different thicknesses under 2 K. As shown in Figure 4, for nanosheets with t>40 nm, we use two-band model to fit the Hall conductivity $\sigma_{xy}$ at 2 K,

$$\sigma_{xy}(2\text{K}) = \left[ n_1\mu_1^2 \frac{1}{1+(\mu_1 B)^2} + n_2\mu_2^2 \frac{1}{1+(\mu_2 B)^2} \right] eB, \tag{1}$$

where $n_1$, $n_2$ are the carrier density in two different bands, and $\mu_1$ and $\mu_2$ are the mobility, correspondingly. Fig. 4A shows the fitting result of Hall conductivity for different nanosheets above 40 nm. The evolution of carrier density as the function of thickness is demonstrated in Fig. 4B. As we can see, in thick nanosheets with t>80 nm, it contains a Dirac electron pocket with high mobility and a secondary electron band with low mobility. This is also consistent with recent ARPES experiments, where an off-centered secondary band is revealed, besides the Dirac band at the Brillouin Zone center (1, 3, 11). The carriers density in both electron pockets decrease as decreasing the thickness, hinting a downward shifting of the Fermi level. Decreasing the thickness further from 80 nm to 40 nm, the carriers in secondary band transforms from n-type to p-type. While below 40 nm, only a single band with p-type carriers is revealed, due to the linear behavior of Hall resistivity versus field. Also, as we can see in Fig. 4B, this p-type carrier density increases further as decreasing the thickness down to 10 nm, demonstrating that, from 160 nm to 10 nm, the Fermi level can be shifted consecutively from conduction band down to the valence band. Our experiment reveals a highly thickness-tuned band topology in ZrTe$_5$ nanosheets.

To gain deeper understanding of the electronic nature of ZrTe$_5$ nanosheets, we show in Fig. 5A the longitudinal magnetoresistance (MR) of the nanosheets with different thicknesses. It is clearly seen that in thick nanosheets (t ≥ 80 nm), the MR presents a quasi-linear behavior with an exponent α close to 1 if simply fitting the MR curve with a formula of $R \propto B^\alpha$. This is well-known for a topological insulator and topological semimetal (27-30), such as in Cd$_3$As$_2$ (28, 30). While for 55 and 40 nm nanosheets, the MR follows a $\sqrt{B}$ type behavior with the fitting parameter α closing to 0.5, indicating the weak antilocalization (WAL) effect in a 3D topological semimetal (31). In contrast, the MR exhibits a simple quasi-quadratic behavior with a fitting parameter α~1.6 − 1.7 as the thickness is below about 40 nm. Interestingly, if we carefully track the MR behavior in very low field region, the MR exhibits a quasi-quadratic behavior with fitting parameter α ~ 2 for thick nanosheets

(t>40 nm), as shown in the log-log plot of the MR curves in Fig. 5B. While for nanosheets below 40 nm, the low field MR shows a $\sqrt{B}$ tendency with α ~ 0.5. As discussed above, the Fermi level shifts downward consecutively to the Dirac point as decreasing the thickness. Thus, the $\sqrt{B}$-type MR induced by WAL effect in the presence of the weak intervalley scattering will surpass the $B^2$-type MR when decreasing the thickness below 40 nm. Unfortunately, we cannot determine whether there is a band gap or not in the Dirac band in the thinner nanosheets, because there is no SdH quantum oscillations in our studied field range. A fact that the nanosheets with t<40 nm show excellent metallic behavior with hole carriers demonstrates that the secondary hole band would dominate the transport and most likely have a conventional nature.

In conclusion, we have systematically studied the transport properties of 3D Dirac semimetal ZrTe$_5$ under different thicknesses. It is found that the resistivity anomaly temperature evolves non-monotonically as decreasing the thickness. By detailed Hall measurements, we find that the the resistance anomalous peak in ZrTe$_5$ nanosheets below 40 nm stems from a totally different mechanism, compared to that above 40 nm. Further analysis of the carrier density demonstrates that the Fermi level shifts consecutively from conduction band to valence band as decreasing the thickness. Our experiments provide a comprehensive insight on thickness-dependent electronic structures in 3D Dirac semimetal ZrTe$_5$, which would be helpful for the understanding of this complex layered material and further study on nanodevices.

**Acknowledgments**

This work was supported by the Natural Science Foundation of China (Grants No. 11374302, No. 11574320, No. 11204312, No. U1432251 and No.U1332139), the program of Users with Excellence, the Hefei Science Center of CAS, the CAS/SAFEA international partnership program for creative research teams of China, and National Science Foundation through grants No. DMR-1004545 and No.DMR-1442366.

**Author contributions**

M. T., W. N. conceived and designed the experiment. X. Z., J. Y., and H. D. fabricated the crystals; J. L., G. Z., X. Z., Y. H., H. Z performed the experiments under magnetic fields. The interpretation was developed by G. Z., J. L., W. N., K. Y, H. L and M. T. The manuscript was prepared by J. L, G. Z., W. N and M. T., together with help of all other co-authors. All authors discussed the results and contributed to the manuscript.

**Figure caption**

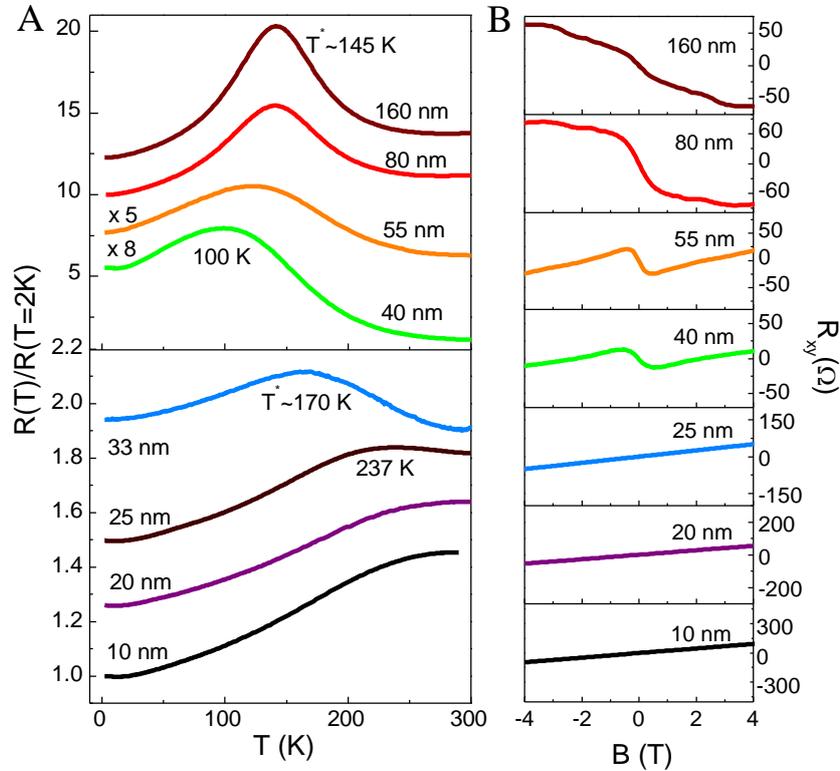

**Fig. 1. Temperature dependent resistivity and Hall resistivity measurements in ZrTe$_5$ nanosheets with different thicknesses.** (A) shows the R-T curves of ZrTe$_5$ nanosheets with thickness above and below 40 nm, respectively. The resistive anomalous peak temperature T$^*$ exhibits a systematic decrease with respect to the thickness down to 40 nm, but then suddenly jumps up with further decreasing of the thickness. All curves are offset vertically for clarity. (B) The corresponding Hall resistivity of the nanosheets, a simple hole carrier band dominates the transport properties in the entire temperature range when t< 40 nm, regardless of temperature being below or above T$^*$.

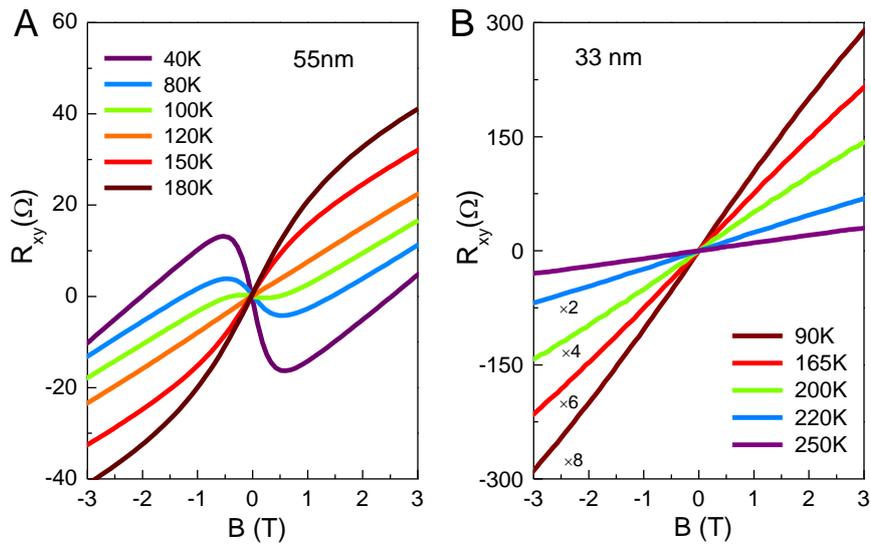

**Fig. 2. The Hall resistivity in ZrTe$_5$ nanoflakes of different thicknesses near the resistive anomalous temperature T$^*$.** (A), (B) show the Hall resistivity in two nanosheets with t= 55 nm and 33 nm, respectively. A typical hole carriers band is found in thinner nanosheets with t<40 nm, regardless of the temperature being below or above the resistive anomalous temperature T$^*$.

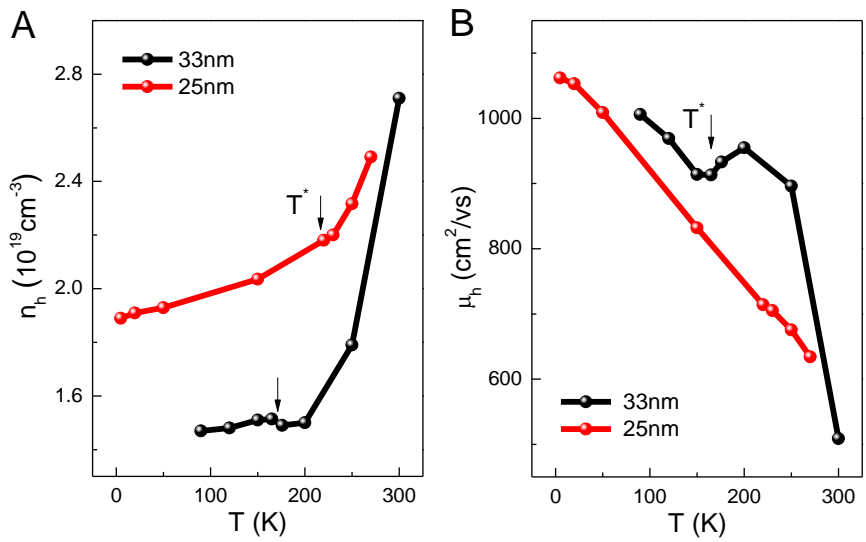

**Fig. 3. The temperature dependence of the carriers density and mobility in 33 and 25 nm thick samples.** (A), (B) Both the carrier density exhibits sharp decrease with temperature down to $T^*$, as marked by the arrows, while the mobility increases correspondingly. The opposite contributions of decreasing charge carrier density n and increasing mobility µ will lead to a resistivity peak near $T^*$.

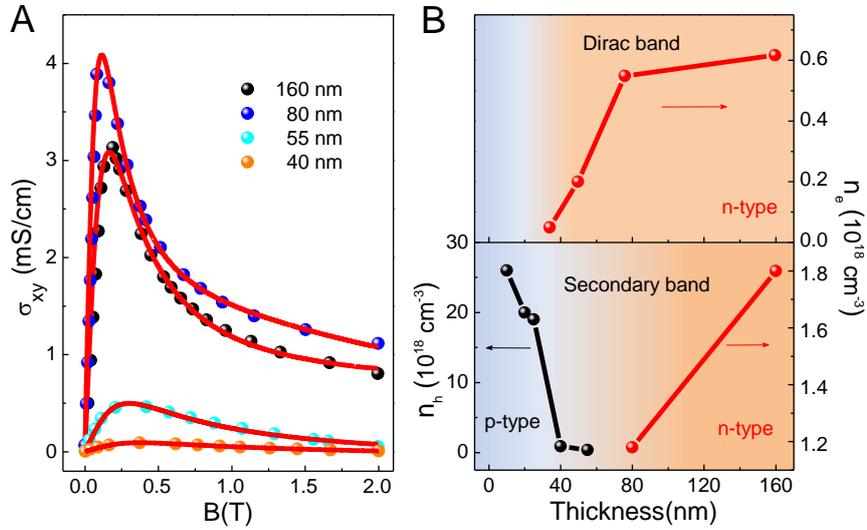

**Fig. 4. The evolution of carrier density reveals a thickness-tuned band topology in ZrTe$_5$ nanosheets.** (A) Two-band fitted Hall conductivity above 40 nm. The red curves are the fitting while the solid circles are the experimental data. (B) The evolution of the carrier density as the function of the thickness. $n_e$ and $n_h$ denote the electron and hole carriers density, respectively. Above 40 nm, the band structure consists of two electron pockets, a 3D Dirac band (with low carrier density and high mobility) and a secondary band. Below 40 nm, only a single hole band is revealed.

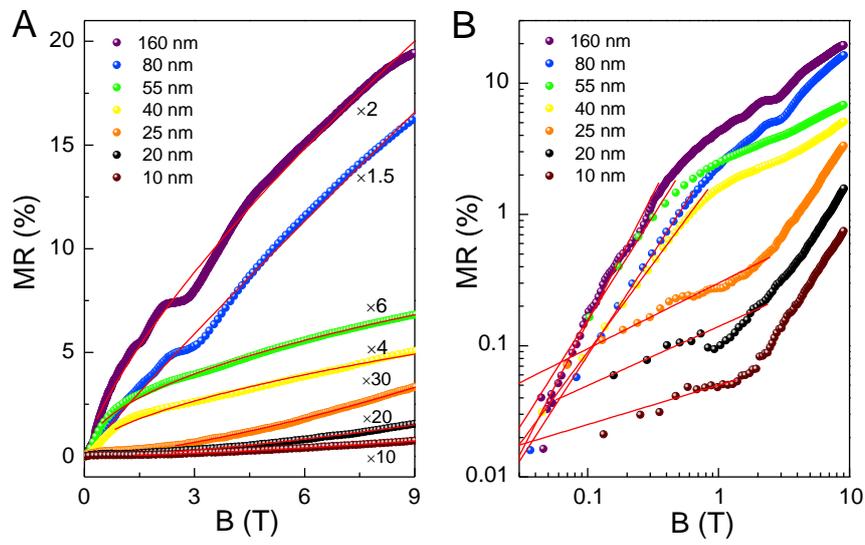

**Fig. 5. The magnetoresistance (MR) curves of ZrTe$_5$ nanosheets with different thicknesses at 2 K**. (A) MR exhibits a quasi-linear or a square-root law of magnetic field above 40 nm. While it transforms to a quadratic law with magnetic field below 40 nm. The red curves are the fitting curves with a formula of $R \propto B^{\alpha}$. (B) shows the MR properties in low field region.